\documentclass[preprint,5p,twocolumn,numbers]{elsarticle}

\usepackage{amssymb}
\usepackage{amsmath}
\usepackage[colorlinks = true,
  linkcolor = blue,
  citecolor = blue,
  urlcolor = blue]{hyperref}
\usepackage{balance}
\usepackage{graphicx}
\usepackage{booktabs}
\usepackage{multirow}
\usepackage{tabularx}
\usepackage{enumitem}
\usepackage{threeparttable}
\usepackage{url}

\begin{document}

\title{A Scenario Analysis of Ethical Issues in Dark Patterns and Their Research}
\author[sdu]{Jukka Ruohonen\corref{cor}}
\ead{juk@mmmi.sdu.dk}
\author[utu]{Jani Koskinen}
\author[sdu]{S\o{}ren Harnow Klausen}
\author[sdu]{Anne Gerdes}
\cortext[cor]{Corresponding author.}
\address[sdu]{University of Southern Denmark, Denmark}
\address[utu]{University of Turku, Finland}

\begin{abstract}
\textit{Context}: Dark patterns are user interface or other software designs
that deceive or manipulate users to do things they would not otherwise do. Even
though dark patterns have been under active research for a long time, including
particularly in computer science but recently also in other fields such as law,
systematic applied ethical assessments have generally received only a little
attention. \textit{Objective}: The present work evaluates ethical concerns in
dark patterns and their research in software engineering and closely associated
disciplines. The evaluation is extended to cover not only dark patterns
themselves but also the research ethics and applied ethics involved in studying,
developing, and deploying them. \textit{Method}: A scenario analysis is used to
evaluate six theoretical dark pattern scenarios. The ethical evaluation is
carried out by focusing on the three main branches of normative ethics;
utilitarianism, deontology, and virtue ethics. In terms of deontology, the
evaluation is framed and restricted to the laws enacted in the European
Union. \textit{Results}: The evaluation results indicate that dark patterns are
not universally morally bad. That said, numerous ethical issues with practical
relevance are demonstrated and elaborated. Some of these may have societal
consequences. \textit{Conclusion}: Dark patterns are ethically problematic but
not always. Therefore, ethical assessments are necessary. The two main
theoretical concepts behind dark patterns, deception and manipulation, lead to
various issues also in research ethics. It can be recommended that dark patterns
should be evaluated on case-by-case basis, considering all of the three main
branches of normative ethics in an evaluation. Analogous points apply
to legal evaluations, especially when considering that the real or perceived
harms caused by dark patterns cover both material and non-material harms to
natural persons.
\end{abstract}


\maketitle

\section{Introduction}

Dark patterns try to exploit people's cognitive biases by deceiving or
manipulating them to do things they would not otherwise do. They are typically
deceptive or manipulative user interface designs, and they usually do the
exploitation for commercial reasons, although many other motives may be present
too. Dark patterns are a thriving research domain in multiple disciplines, but
there are still gaps in the literature to fill. This paper fills one such
gap. To elaborate where the gap specifically is, the paper's background and
framing should be briefly discussed.

The paper contributes to the desirable but still limited efforts to build a
systematic ethics toolkit in software engineering \citep{Aydemir18} by providing
new material on ethics statements in a particular, previously unexplored,
relevant, and timely context. Although initial ethical assessments in software
engineering trace to the early 1990s~\citep{Gottenbarn91}, if not earlier,
ethics have still received relatively limited attention in the discipline,
despite frequent calls for better ethical engagement~\cite{Rashid09}, including
in terms of empirical research practices~\cite{Hall01, Singer02, Vinson08}. The
situation seems to be similar in nearby disciplines, such as information systems
research~\cite{Hassan2018, Rogerson2019, Clarke20}, the exception of artificial
intelligence (AI) ethics perhaps notwithstanding, which still has its own
challenges \cite{Bleher23, Mittelstadt19, Resseguier20}. Despite these
limitations, ethical questions are important throughout computing
disciplines. Ethics examine normative questions about people's morally right or
wrong behavior, and many such questions are closely related to values people
have.

Traditionally, the concept of value in software engineering has been defined in
terms of economics; monetary costs and revenues have been the main
focus~\cite{Boehm06}. Driven by concepts such as responsible software
engineering~\cite{Schieferdecker20}, design responsibility~\cite{Gray18}, value
sensitive design~\cite{Gerdes23}, and values conscious
software~\cite{Hussain20}, there has recently been an increasing interest to
examine how also other values, including those motivated by ethics and those
wanted by a society, can be incorporated into software and its
requirements~\cite{Levina24, Perera19}. This value-motivated software
engineering research has also examined values held by software engineers working
in the software industry~\cite{Hussain20, Winter18}. Against this background,
among other things, it has been asked what is a moral character of a user
interface designer who designs dark patterns~\cite{Gray18}. While an answer to
this question is still open, and likely continues to be so also in the future,
the question is a good example about applied ethics. In fact, an uncompromising
person might say that a software engineer involved in designing of dark patterns
has a questionable moral character because exploiting people is morally wrong. A
slightly less uncompromising person might agree but he or she might point out
that the question is also about responsible software engineers who comply with
the discipline's and the software industry's professional ethics. Despite such
compliance, it should be understood that a design can be ethically questionable
for reasons other than those related to virtue ethics. For this reason, the
paper considers also potential violations in terms of consequentialist
(utilitiarianist) and deontological ethics.

The question and associated controversies surrounding dark patterns are also
about law. In this regard, the paper also contributes to the nascent but rapidly
increasing literature on the regulation of dark patterns~\cite{Brenncke24,
  DiPorto24, Herman24, Trzaskowski24}. Though, the legal analysis pursued is
intentionally kept minimal and plain; legal aspects are only discussed to the
extent necessary for the applied ethics evaluation. To simplify the legal
discussion and narrow the paper's scope, the discussion is also restricted to
laws enacted and enforced in the European Union (EU). As will be seen, these EU
laws build upon the two principal theoretical concepts behind dark patterns,
deception and manipulation.

What makes the paper unique is that the ethics of dark patterns and their
designing are explicitly connected to research ethics. Through this connecting,
the paper is able to discuss with and contribute to the noted general literature
on ethical practices in software engineering research. To this end, the focus is
more on the ethical problems that may arise when doing research on dark
patterns. Even when one would take a hard stance and perceive all dark patterns
as morally wrong, a question would still remain about the ethics of studying
morally wrong things. More practically, the core concepts involved, including
deception in particular, are well-suited for approaching difficult ethical
problems in scientific software engineering research projects. A further
important point is that the paper extends the concept of dark patterns to cover
also other deceptive or manipulative software solutions. Cyber security would be
a good example in this regard. As has been pointed out in existing research
studying people's ability to detect dark patterns, many dark patterns are
closely related to phishing and its techniques~\cite{BongardBlanchy21}. In fact,
it could be argued that replicating a bank's website and then trying to
manipulate people to use it would be a dark pattern, or a so-called dark
strategy~\cite{Bosch16}, \textit{par excellence}. Cyber security is a good
example also in the sense that not all deception techniques therein are
necessarily morally wrong.

The paper's background is discussed in more detail in Section~\ref{sec:
  background}. The discussion proceeds from an elaboration of the research on
dark patterns, deception, and manipulation to the EU laws considered. Then,
Section~\ref{sec: methods} presents the methodology used. In particular, it
introduces the analytical framework for the ethical analysis of theoretical
scenarios. The framework is based on the three main branches of normative
ethics, hence offering applicable ethical tools for software
  practitioners and researchers to deal with dark patterns. The subsequent
Section~\ref{sec: scenarios} present the scenarios, each of which is carefully
contextualized around software engineering research projects. To raise the
practical relevance, many of the scenarios presented are specifically about
university-industry projects. The final Section~\ref{sec: discussion} presents a
conclusion, a brief take on limitations, and a discussion about further research
potential paved by the paper.

\section{Background}\label{sec: background}

\subsection{Dark Patterns}

Dark patterns are user interface designs or other software solutions that
covertly or overtly make people do things they would not otherwise do. Some dark
patterns deceive people, while others manipulate them~\cite{Narayanan20}. As
soon discussed in detail, deception and manipulation are the core concepts also
for the present work. It is also important to start by emphasizing that the
opening sentence providing a definition extends the concept of dark patterns
beyond user interface designs. Recommender systems would be a good example in
this regard, especially since many Europeans perceive those as
manipulative~\cite{Ruohonen24FM}. However, people's perceptions do not
necessarily equate to actual manipulation, which correlates with technical
functionalities, as discussed later in Section~\ref{subsec:
  regulations}. Another point is that not all dark patterns are necessarily
about deception or manipulation, although these are the core concepts adopted
for the present work for brevity.

The earliest online origins trace to 2010 when dark patterns were first used as
a concept to warn about tricks used by online marketplaces to make people buy
unwanted things~\cite{Herman24, Trzaskowski24}. Since then, these patterns have
become extremely common in online market places according to empirical research
results~\cite{Mathur19}. In addition, they are used also on various other online
media platforms, including social media. Also awareness has increased. In fact,
there is even an online catalog for common dark
patterns.\footnote{~\url{https://hallofshame.design/collection/}} In addition to
user interface design, the general background relates to the psychology of
cognitive biases, behavioral economics, so-called nudging, A/B testing,
attention seeking and celebrities, and increasingly commonplace digital
addiction~\cite{Clarke20, Narayanan20}. Here, the popular jargon term nudge
generally refers to any attempt that seeks to influence people's behavior
through their cognition but without limiting their freedom of
choice~\cite{Kuyer23}. Given this background, it is also understandable that
legally dark patterns are closely related to laws on deceptive marketing
practices and consumer protection jurisprudence in general. That said, it
remains debatable whether nudging can be equated to manipulation, as it
proponents would probably deny; in any case, it is not illegal.

The overall building block behind dark patterns is the extensive scientific
research on human decision-making and its caveats. The basic lesson from this
research is that people often make bad decisions due to limits in their
cognitive capabilities. Dark patterns exploit particularly real or perceived
cognitive biases people have. Numerous such biases have been identified and
examined in relation to dark patterns~\cite{Drumwright15, Mathur19}. A good
example would be so-called scarcity bias; people tend to place more value on
things they believe are scarce. A corresponding dark pattern would then present
something as scarce, potentially even in case it is not scarce but
plentiful. For instance, an online platform for reserving hotel rooms might
deceive people by fooling them to believe that most hotel rooms in a city have
been already reserved, although in reality plenty of rooms would be
available. Another common example would be to exploit the time available for
purchasing decision-making by falsely presenting urgency in terms of a countdown
timer or other user interface design pattern. While these cases are relatively
straightforward to deduce about because they are all related to real or
potential financial harms to consumers, there are numerous other dark patterns
that enlarge the scope to non-material harms. Such harms make legal and ethical
interpretation challenging because the scope is extended toward questions
involving loss of autonomy and control, labor and cognitive burdens, emotional
distress, and privacy violations, among other things~\cite{Gunawan22}. Such real
or potential harms make it understandable why the research and legal practice on
dark patterns have both taken a generally strongly negative standpoint regarding
the patterns.

The negativity is already embedded to the whole term \textit{dark} patterns. By
following recent research~\cite{ZhangKenney24}, the present paper considers also
so-called \textit{gray} patterns and generally the ethical questions that arise
when operating in a gray area of user interface design and software engineering
research in general. In addition, the paper extends the concepts of deception
and manipulation toward research ethics and research integrity. In other words,
the paper's focus is not only on dark patterns but also on the design and
development of these, albeit only in terms of theoretical, imaginary
scenarios. This wider scope raises also the paper's practical relevance for
software engineering both as a scientific discipline and as an industry
practice.

\subsection{Deception}\label{subsec: deception}

For the present purposes, deception can be defined as a way ``\textit{to
  intentionally cause another person to acquire or continue to have a false
  belief, or to be prevented from acquiring or cease to have a true
  belief}''~\cite{Pawlick19}. Although making someone to have a false belief or
to prevent him or her to have a true belief often connotes with lying, which
many people perhaps perceive as a morally bad behavior, the ethics of lying are
a classical but not a straightforward philosophical
topic~\cite{Isenberg64}. While hardliners would say that all lying is bad and
thus unethical, there are also many other things to
consider~\cite{Noggle22}. Among other things, to pursue this line of reasoning
further, one would need to also consider plain lying and other non-lying
deception separately, although many would still end up with the conclusion of
unethical behavior~\cite{Gillon93}. However, from a perspective of utilitarian
ethics, a concept soon elaborated in Section~\ref{subsec: utilitarianism}, many
applied cases can be postulated in which deception can be seen as ethical. It
should be also understood that numerous deception techniques are used even by
ordinary people in their daily lives, some perhaps without them being aware of
their deceptive nature. Even women who use makeup could be seen as using
deception.

Military ethics would be a good example. Therein, deception connotes with many
military techniques, such as camouflage, which may save lives of soldiers and
civilians, and would thus be oftentimes (but not always) seen as
ethical~\cite{Rothstein13}. Cyber security would be another good
example. Therein, many defensive cyber security solutions build upon deception
that can be either ethical or unethical depending on a context and a given
deception technique~\cite{Pawlick19, Rowe09}. Insider threats and social
engineering would constitute a good example already because both are often about
human weaknesses. In general, in this setting external threat actors may lure
insiders in an organization to compromise the organization by deceiving them and
manipulating their psychological weaknesses~\cite{Ruohonen24ICDF2C}. Although
compromising an organization is a cyber crime and thus rather unambiguously
unethical, there are also many other scenarios in which social engineering of
insiders may be seen as ethical, including cases whereby an organization in
question or the ethical third-parties it has contracted try to improve the
organization's cyber security posture by raising security awareness among
staff~\cite{MoutonKimppa15}. As is typical in ethics, these examples demonstrate
that not a single one-size-fits-all answer exist for determining whether
deception is morally right or wrong.

The ethics of deception have been examined also in research settings whereby
some authors have argued that deception can be ethically used to obtain human
subject data for research, especially when done in conjunction with a
debriefing~\cite{Goode96, Resnik18, Smith83}. Rather similar arguments have been
raised in terms of human subject data collection through covert observing of
people; here, to obtain consents of the people observed later on, their
anonymity and trust in the now overt observers have been argued to be
important~\cite{Podschuwei21}. These points notwithstanding, the well-known and
high-profile legal and ethical scandals, such as the Cambridge Analytica one,
have made straightforward argumentation less convincing in the information
technology domain. In other words, deception may be ethical in some
  research settings, but other ethical lapses may also occur.

Against this backdrop, the scenarios evaluated consider the science's
\textit{de~facto} practice of obtaining informed consents from human study
subjects. This practice is in a contrast with deception, which, in the context
of human subject research, means that the true intent and methods of a study are
not revealed to study subjects beforehand~\cite{Vinson08}. Then: although there
are no universally accepted definitions for informed consent, it is usually seen
to contain a full and detailed disclosure about a given research project, a
comprehension and competence requirement of the human study subjects, an option
for them to withdraw from the research at any time, and their actual, voluntary
consents to take part in the research project~\cite{Singer02, Vinson08}. A lot
of research has been done to examine these criteria, including in terms of how
to ensure the comprehension requirement through written consent
forms~\cite{Schenker10}. As only analytical scenarios are considered, the
assumption in this paper is that the criteria noted have been fully satisfied.

In addition, to increase the analytical depth and practical relevance of the
scenario analysis, consenting is also considered in terms of data protection
regulations, including the General Data Protection Regulation (GDPR) in
particular.\footnote{~Article 6(a) in Regulation (EU) 2016/679.} That said, it
should be understood that a consent is not the only way to obtain a legal basis
for personal data processing; a scientific research project may also rely on a
clause about public interests.\footnote{~Article~6(e) in Regulation (EU)
  2016/679.} In some fields, such as clinical medicine, there may also be legal
obligations to archive research data for a long period of
time.\footnote{~Article 6(c) in Regulation (EU) 2016/679.} A further relevant
point is that the GDPR's consent has also been frequently criticized; technical
violations are common and even without violations people seldom understand to
what they are consenting to when using digital platforms, applications, and
services~\cite{Nguyen22, Ruohonen23DS}. Despite this criticism, it assumed that
also the data protection consent requirements have been
fulfilled.\footnote{~Article 7 in Regulation (EU) 2016/679.} The same
applies to other legal bases, which are always assumed to be valid. The final
point is that in some cases a separate consent may be required even in case a
consent is not used as the primary legal basis for processing. These cases
relate to sensitive data, such as those revealing ethnic origin, political
beliefs, or sexual orientation.\footnote{~Article 9(1) in Regulation (EU)
  2016/679.} Although detailed legal interpretations can be
difficult~~\cite{ChenDove22}, it is again also possible to rely on an exemption
for scientific research.\footnote{~Article 9(2)(j) in Regulation (EU) 2016/679.}
Regardless of a particular legal basis, again the validity of a basis is assumed
during start of a project. This choice simplifies the analysis. While validity
is assumed initially, the interest is to examine different scenarios that
involve legal uncertainties due to changes in a project's life cycle.

\subsection{Manipulation}\label{subsec: manipulation}

Then there is the concept of manipulation. Given the paper's framing toward
algorithmic manipulation, it can be understood in the present context as an
intentional algorithmic attempt to change a person's behavior. If a manipulation
is successful, the person does something he or she would not otherwise
do. Although the definition used is simple, it is also problematic because it
overlaps with other concepts, such as those used in the field of persuasive
technologies~\cite{Fogg08}. In other words, the definition covers anything and
everything from persuasion to harder techniques, starting from pressuring and
ending to coercion.

However, the definition does not necessarily involve an attempt to change a
person's behavior against his or her will, which has been a common theoretical
standpoint in the literature~\cite{Gray21, Jongepier22}. Another point is that
many algorithmic manipulation attempts are covert in their aims, but oftentimes
manipulation is overt in practice, meaning that a person is fully aware of
it. According to surveys and other assessments, many people indeed are fully
aware that they are being manipulated online by user interfaces and
algorithms~\cite{BongardBlanchy21, Gray21, Ruohonen24FM}. These and other
examples underline that covert or hidden influence is not a necessary condition
for manipulation~\cite{Jongepier22}. Unlike the more positive nudging
  concept, the definition also allow limiting people's freedom of choice by
restring the available choices for decision-making~\cite{Schmidt20,
    Weinmann16, Wilkinson12}. It can be further noted a manipulation may
involve also deception in which case a successful manipulation would change a
person's behavior through changing his or her beliefs. That said, manipulation
is commonly seen to involve also emotions and desires. Due to these reasons, the
scenario analysis does not separate the two; behaviors and beliefs, including
behavioral beliefs, are taken together.

Although the context is algorithms, the notions from conventional,
inter-personal manipulation apply well; a system that controls given algorithms
is assumed to have some knowledge about a person with which it makes predictions
and inferences about the person's intentions with regard to specified behavioral
acts or outcomes~\cite{Kligman92}. This kind of a manipulation is thus usually
probabilistic~\cite{Christiano22}. In fact, the whole A/B testing paradigm
relies on probabilistic reasoning. As knowledge about a person implies existence
of personal data about the person, also the GDPR is involved. Yet, in terms of
ethics, it seems clear that merely attempting to change a person's behavior, or
successfully doing so, cannot count as a criterion for determining whether the
associated act is ethical or not. Otherwise, even democracy would be doomed
because every electoral campaign would be seen as unethical, among other
things. Therefore, a utilitarian analysis would focus on whether the specified
outcomes of behavioral change are ethical.

Another option would be to continue the lying theme and consider unethical
manipulation to involve bypassing, circumventing, or otherwise undercutting a
person's rational capabilities, thus also reducing his or her personal autonomy
and free will~\cite{Christiano22, Noggle22}. Autonomy and free will are
fundamental already because they are necessary but not sufficient conditions for
ethical reasoning to begin with, especially in deontology. However,
there are many philosophical demarcation problems between manipulation,
autonomy, and rationality~\cite{Jongepier22}. The emphasis on rationality might
also be a bad choice in general because it would lead to the long-lasting debate
over whether consumers or even citizens in general are rational to begin
with. If it is accepted that many, if not most, people carry out decisions based
on intuitions, desires, and feelings instead of a rational calculus, as
suggested in the literature~\cite{Drumwright15, Narayanan20, Noggle22}, then it
also becomes difficult to use interfering with a person's rationality as a
criterion to determine ethics of manipulation; after all, rationality is not a
decisive factor to begin~with.

A further option would be to again focus on consent~\cite{Sunstein16}. If a
person voluntarily makes an informed decision to join something he or she
believes that will change his or her behavior through overt manipulation, the
manipulation would be consequently seen as ethical. Many people indeed do such
decisions, whether the context is about losing weight, stopping smoking, or
learning software engineering. However, this option is open to the same
criticism noted already about the GDPR's consent and, perhaps to a lesser
extent, informed consent in science. In other words, it is open to a debate
whether people really understand all the elements involved in algorithmic
manipulation and its real or potential consequences. In terms of science and
research, the ethical problems resemble those encountered in action research,
which too attempts to make changes through interventions~\cite{Williamson02}. As
will be elaborated in the next Section~\ref{subsec: regulations}, it is also
possible that dark patterns are specifically used for manipulating consenting
itself~\cite{Gray21, Gunawan22}. Furthermore, even under informed consent,
neither people nor software engineers of dark patterns may fully understand the
consequences, including potential harms caused. Digital addiction would be a
good example in this regard.

A final common approach to deduce about the ethics of algorithmic manipulation
involves contemplating about its larger, societal consequences. In particular,
algorithmic manipulation is widely perceived to interfere with some of the core
pillars of democracy~\cite{Christiano22, Ruohonen24FM, Sunstein16}. In general,
democracy and political legitimacy are seen by many political scientists to
require an informed public that observes, understands, and deliberates upon
politics and policies. To this end, a lot of criticism has been expressed that
algorithmic manipulation, including nudging, distorts the principles of
deliberation and public discourse~\cite{Kuyer23}. Allegedly algorithmically
driven polarization would be the prime example in this regard. Although many of
the associated topics and consequences are unclear and actively
debated~\cite{Stark20}, and democracy in itself seems a far-fetched topic to
consider in software engineering as a scientific discipline, the societal
consequences align well with the noted concept of responsible software
engineering. The scenarios presented also involve a case that showcases how
ethical dilemmas related to democratic principles can arise in practice also in
software engineering.

\subsection{Regulations in Europe}\label{subsec: regulations}

The European Union has a long history of either implicitly or, more recently,
explicitly regulating dark patterns. In fact, there are at least seven EU laws
addressing these. By implication, it is not a surprise that criticism has been
expressed about regulatory redundancy, overlap, inconsistency, and
fragmentation~\cite{Herman24}. In what follows, the seven laws are briefly
addressed. The focus of the concise review is on the definitions.

First, as was already noted, algorithmic manipulation involves personal data,
and thus the GDPR applies. The fundamental data protection concept of
informational self-determination can be also related to the concept of
individual autonomy~\cite{Trzaskowski24}. The supervising authorities have also
released guidelines for dark patterns, emphasizing particularly their relation
to fairness, informed consent, and transparency of personal data
processing~\cite{EDPB22}.\footnote{~Articles 5, 7, and 12 in Regulation (EU)
  2016/679.}  Besides the collection and processing of personal data, it is
worthwhile to remark the GDPR's provision for opting out from automated
decision-making and profiling, although the provision's legal interpretation is
debated~\cite{Ruohonen24FM}. Then, second, there is a directive concerning
generally unfair business-to-consumer practices. It defines a misleading
commercial practice as a one that ``\textit{contains false information and is
  therefore untruthful or in any way, including overall presentation, deceives
  or is likely to deceive the average consumer}''.\footnote{~Quoted from Article
  6(1) in Directive 2005/29/EC.} Thus, the definition builds upon deception but
not manipulation. Third, there was a directive for financial services that also
implicitly dealt with dark patterns.\footnote{~Directive 2002/65/EC} Although it
has been discussed in the literature at length~\cite{Brenncke24, Herman24}, it
was recently repealed and thus not worth considering in detail any
more.\footnote{~Directive (EU) 2023/2673.}

From the more recent regulations, fourth, the so-called Data Act also implicitly
regulates dark patterns. In a context involving users sending data to
third-parties, this regulation prohibits making the exercising of choices and
rights of users ``\textit{unduly difficult, including by offering choices to the
  user in a non-neutral manner, or by coercing, deceiving or manipulating the
  user, or by subverting or impairing the autonomy, decision-making or choices
  of the user}''.\footnote{~Quoted from Article 6(2)(a) in Regulation (EU)
2023/2854.}  In addition to deception, this regulation considers also
algorithmic manipulation and even talks about harsher measures such as
coercion. However, a definition is lacking for manipulation, which is a problem
in all of the EU's laws implicitly or explicitly dealing with dark
patterns~\cite{Brenncke24}. Fifth, also the so-called Digital Markets Act (DMA)
implicitly regulates dark patterns. In particular, so-called large gatekeeper
platforms should not design their user interfaces ``\textit{in a way that
  deceives, manipulates or otherwise materially distorts or impairs the ability
  of end users to freely give consent}''.\footnote{~Quoted from Recital 37 in
Regulation (EU) 2022/1925.} Once again, both keywords, deception and
manipulation, are present. Sixth, also the so-called AI Act implicitly addresses
dark patterns. In particular, such an AI system is prohibited that
``\textit{deploys subliminal techniques beyond a person’s consciousness or
  purposefully manipulative or deceptive techniques, with the objective, or the
  effect of materially distorting the behaviour of a person or a group of
  persons by appreciably impairing their ability to make an informed decision,
  thereby causing them to take a decision that they would not have otherwise
  taken in a manner that causes or is reasonably likely to cause that person,
  another person or group of persons significant harm}''\footnote{~Quoted from
Article~5(1)(a) in Regulation (EU) 2024/1689.}  While it is interesting that the
regulation envisions subliminal AI techniques, the basic building blocks are
again present: manipulation, deception, and their effect upon
decision-making. What separates this AI regulation from the others is that it
explicitly restricts the prohibition only to those systems that cause or are
likely to cause significant harm.

Seventh, particularly relevant for the present purposes is the so-called Digital
Services Act (DSA) enacted in 2022 and enforceable in 2024. Although most of the
laws noted are related, the DSA is worth taking under a closer look. The DSA
imposes various requirements for online platforms and services, including upon
content moderation~\cite{Eder24} and recommender systems
\cite{Ruohonen24FM}. Although a particular focus is on very large platforms and
search engines, requirements are imposed also upon smaller platforms and digital
services in general. Then, to continue to the actual topic at hand, the DSA in
fact explicitly forbids dark patterns: those who develop, provide, and maintain
``\textit{online platforms shall not design, organise or operate their online
  interfaces in a way that deceives or manipulates the recipients of their
  service or in a way that otherwise materially distorts or impairs the ability
  of the recipients of their service to make free and informed
  decisions}''.\footnote{~Quoted from Article 25(1) in Regulation (EU)
  2022/2065.} This prohibitive definition resembles the one in the Data Act and
the consent-specific clarification in the DMA.

There are several important remarks to be made about this explicit prohibition
of dark patterns. First, it is remarkable that the prohibition applies not only
to deployments but even to the design of dark patterns in the context of online
platforms. However, as designing is typically an internal matter of software
development companies, enforcement may be difficult but not
impossible. Nevertheless, the coverage of designing itself aligns with the
concept of anticipatory ethics soon discussed in Section~\ref{subsec: scenario
  analysis}, further aligning with the Collingridge's classical dilemma; it is
difficult---if not impossible---to robustly predict an emerging technology's
future social consequences~\cite{Collingridge90}. Second, the prohibition
applies to most so-called closed and open platforms. In general, a platform's
openness depends on various factors~\cite{deSouza16}, but in the present context
it can be understood to depend upon whether a user of a platform needs to
register and login to the platform before its use. Closed platforms require such
a registration and a login, whereas open platforms are open for anyone to view
content. However, the regulation clarifies that only those closed platforms are
covered whose registration is open for everyone, meaning that those closed
platforms are excluded whose registration involves a human-based vetting
procedure.\footnote{~Recital 14 in Regulation (EU) 2022/2065.} This point is
important for the scenarios considered.

The notion about viewing of content is important because an online platform is
defined to mean ``\textit{a hosting service that, at the request of a recipient
  of the service, stores and disseminates information to the
  public}''.\footnote{~Quoted from Article 3(i) in Regulation (EU) 2022/2065.}
The definition continues to exclude platforms whose activity is only minor or
whose features are purely ancillary functions of another service. Then,
recipients are defined as both natural and legal persons who use a platform
particularly for the purposes of seeking or disseminating
information.\footnote{~Article 3(b) in Regulation (EU) 2022/2065.} Finally, to
disseminate to the public means ``\textit{making information available, at the
  request of the recipient of the service who provided the information, to a
  potentially unlimited number of third parties}''.\footnote{~Quoted from
  Article 3(k) in Regulation (EU) 2022/2065.}

The third and fourth remarks follow: the exclusion of minor information
dissemination is important but no explicit definition is given beyond the notion
about potentially unlimited third-parties, and the definition covers both
commercial for-profit platforms and not-for-profit platforms.\footnote{~Though,
  recital 13 in Regulation (EU) 2022/2065 gives an example about the comments
  section in an online newspaper as a mere ancillary function. It also notes
  that cloud computing and web hosting services are excluded. Recital 14 further
  notes that so-called interpersonal communication services, as defined in
  Directive (EU) 2018/1972, are excluded; these include emails and mobile
  communication applications insofar as they do not involve dissemination to a
  potentially unlimited number of recipients. Despite these clarifications, the
  definition for a minor storing and dissemination of information is unclear.}
Again, these points are important for the ethics scenarios presented. The fifth
and last point is fundamental: the definition of a dark pattern is based
explicitly on the concepts of deception and manipulation, and both are framed
against algorithms that distorts or otherwise impairs the ability of a recipient
to make free and informed decisions. As was noted in Section~\ref{subsec:
  manipulation}, the notions of free will and informed decision-making are
difficult concepts theoretically---and legally. A general interpretation is that
the DSA's definition emphasizes individual autonomy as a normative
characterization, combining both individual decision-making and the options upon
which decisions are made~\cite{Brenncke24}. On one hand, the latter, the options
available for decision-making, may be fairly easy to interpret legally; for
instance, a platform might unfairly restrict the available options for some or
all people, which would be relatively straightforward to deduce about. On the
other hand, there is a delicate but challenging debate in the literature about
individual autonomy, law, and ethics~\cite{Gunawan22, Kuyer23}. Although
rationality does not appear in the DSA's definition, the notion of informed
decision-making is likewise much more difficult to reason about.

If people make decisions based on emotions and intuitions, can the decisions
made really be informed? A similar question can be asked with respect to
informed consent in science and the GPDR's consent, which too should be
informed. However, a legal interpretation may not need to answer to such
difficult questions, but it may instead rely on the definition of deception and
particularly its notion of causing someone to hold false
beliefs~\cite{Brenncke24}. This framing to deception would presumably allow to
legally intervene in various dark patterns seen in the wild, including deceptive
online banners for the GDPR's consent, price manipulation through falsely
equating headline prices to final prices, and so-called fake scarcity via which
a consumer is deceived to believe that only a few items are in a stock. All of
these examples are fairly objective facts about a deceiving platform, and hence
a legal system may not necessarily need to answer to a question whether they
actually manage to successfully deceive people. At least an intention to deceive
or manipulate can be ruled out legally~\cite{Trzaskowski24}. In other words,
what particularly matters in this line of reasoning is the method, the technical
functionality~\cite{DiPorto24}. As online deception may be textual, visual, or
something else~\cite{Rowe09}, it is also worthwhile to note that the DSA covers
all deception techniques in terms of technology.\footnote{~Recital 67 in
  Regulation (EU) 2022/2065 makes this clear by noting visual, auditory, and
  other components for deception.}

Finally, it is worth returning to the argument that algorithmic manipulation may
be particularly relevant for democracy and democratic deliberation. Here, the
algorithmic amplification of disinformation would be a good example. In general,
the DSA imposes various mandates for large platforms in this regard, including
upon notification and takedown measures, advertising transparency, risk
assessments, due diligence, and auditing. Due to the complexity of the mandates
and the fundamental nature of the problem, including its relation to the freedom
of expression, criticism has been common and the effectiveness of sound
enforcement remains an open question~\cite{Nannini24, Ruohonen24FM}. However,
what matters for the present purposes is that there is a discrepancy in the DSA
between algorithmic manipulation as a dark pattern technique and algorithmic
manipulation as a way to disseminate and amplify disinformation. Regarding the
latter, the examples include intentional manipulation or deception of a platform
and its recommender system by bots or fake accounts, but no connection is made
in the regulation to a question on how such a manipulation is related to a
user's ability to make free and informed decisions based on the manipulated
information. In this case there are two manipulators: the party who is
controlling the bots or fake accounts and the given platform whose algorithms
allow such behavior, potentially making the situation worse by unintentional
amplification of bad content. While a legal or other interpretation to this
problem can be left for further work, it would seem reasonable to initially
deduce that the DSA's definition for a dark pattern might, or perhaps should,
apply also in this disinformation-related manipulation context.

\section{Methods}\label{sec: methods}

\subsection{Ethical Evaluation}\label{sec: ethical evaluation}

Broadly speaking, ethics as a discipline can be framed with three broad
categories: meta-ethics, normative ethics, and applied
ethics~\cite{Wright14}. The first considers questions such as the meaning of
moral discourse, metaphysics, epistemology, and objectivity of moral
questions~\cite{Miller2003}. Although beyond the paper's scope, meta-ethics
establish the philosophical basis for normative theories and their underlying
assumptions \cite{Stahl12}. Then, normative ethics deal with moral standards for
right and wrong. There are three main normative ethical branches:
consequentialism, deontology, and virtue ethics. These different normative
approaches offer different philosophical justifications for defining what is
ethical. Finally, applied ethics focus on specific moral issues in a particular
context~\citep{Wright14}. It is the cornerstone for approaching ethical
questions in software engineering. In practice, the three normative branches are
often used for ethical evaluations in applied contexts~\cite{Isenberg64,
  MoutonKimppa15}. Also the present work builds upon these.

\subsubsection{Deontology}\label{subsec: deontology}

The first branch is deontology. It posits basic \text{rules---duties}
  and observance of rights---that one should follow to act morally. Moreover, in
  deontology, the autonomy and rationality of a person are preconditions for
  determining whether the morality of the person's acts can be evaluated in the
  first place~\cite{Darwall06}. Thus, a person committed to deontological
ethics believes that there are some universal rules, such as human rights, on
what is right or wrong. It is also possible to approach deontological ethics by
philosophical principles, such as Immanuel Kant's famous categorical
imperative according to which one should only commit acts that one would be
willing to become universal laws, and treating people as an end
  themselves, never as mere means or tools only~\cite{Kant70}. Another
deontological principle for applied evaluation purposes has been a
contextual maxim of ``do onto others only that to which they have
consented'' \cite[p.~117]{MoutonKimppa15}. Although not without problems
  in some contexts---as cases such as parenting would readily indicate, this
simple principle seems as a suitable starting point for the present purposes
because it aligns directly with the scientific and data protection requirements
for a consent.

However, as was discussed in Section~\ref{subsec: deception}, there are also
many other ways to legally or otherwise justify a project's data collection and
analysis. Thus, in general, the deontological evaluation is based on a more
general and simplified maxim that what is legal is also ethical. Although this
maxim is subject to a justified long-standing debate, as cases like civil
disobedience demonstrate, it aligns well with the ethical software engineering
principle of accountability~\cite{Rashid09}. Having said that, the evaluation is
not based on an extreme deontological position of ``rules-without-consequences''
but rather follows a more moderate principle of ``rules-with-consequences'',
which has been advocated also for software engineering~\cite{Genova07}. The
principle is rather closely associated with a broader notion of just (rule)
consequentialism, which would further extend an evaluation toward ethics of the
rules themselves and their consequences~\cite{Moor99}. Not only may rules be
ethically wrong---in which case a virtuous person might seek to change them, but
ethics should ideally drive also policy-making and enactment of rules. However,
as the regulations considered have already been enacted, the
``rules-with-consequences'' principle suffices for the evaluation.

In particular, the GDPR is in many ways a risk-based regulation; throughout the
regulation, potential risks to natural persons are emphasized. Risks are also
behind the regulation's notion of data-protection-by-default, which aligns with
the mandate to carry out data protection impact assessments in high-risk
scenarios.\footnote{~Article 25(1) and Article 35(1) in Regulation (EU)
  2016/679.} Although the terminology is somewhat loose, and thus still
evaluated and debated~\cite{Eder24, Efroni21}, rather similar points apply to
the DSA, which is particularly concerned with so-called systemic risks to whole
societies.\footnote{~Although an exact definition is lacking, recitals 80 to 83
  in Regulation (EU) 2022/2065 specify four categories of systemic risks:
  illegal content, threats to fundamental rights, consequences to democratic
  processes, including deliberation and electoral processes, and risks to public
  health, including physical and mental well-being of citizens and minors in
  particular. The risks assessments specified in Article 34 and other related
  requirements only cover so-called very large online platforms and search
  engines.} It is also worth emphasizing that the ``rules-with-consequences''
principle is also explicitly embedded particularly to the GDPR, the DSA, and the
DMA, all of which entail heavy financial penalties or other serious consequences
from non-compliance.

\subsubsection{Utilitarianism}\label{subsec: utilitarianism}

Utilitarianism draws from consequentialism; the rightfulness of an action is
determined by its consequences. In other words, outcomes are what
matters. According to classical utilitarian theorists, such as Jeremy
Bentham~\cite{Bentham24} and John Stuart Mill~\cite{Mill98}, utility was
essentially seen something beyond a quantified measure, such as economic
revenue, but instead related more to concepts such as happiness, pleasure, and
well-being. Such concepts reiterate the point in the introduction about
value-based software engineering.

In practice, ethical evaluation based on utilitarianism can be done by
considering effects upon a single individual or a larger group of
individuals. The latter is followed in the present work. By following existing
research~\cite{MoutonKimppa15}, the evaluation considers the effects upon the
majority of people in a group, an organization, or a society at large. To again
avoid an extreme utilitarian standpoint of ``consequences-without-rules''
\cite{Genova07}, the evaluation is balanced against the deontological and other
evaluations.

In addition to the risks already mentioned, the real or potential consequences
can be considered in terms of scientific research. Here, the avoidance of harm
to human subjects and non-maleficence in general are the fundamental ethical
principle to always follow. Respecting individual autonomy and justice would be
further common principles~\cite{Whittlestone19}. In addition, confidentiality,
beneficence, and scientific value can be
considered~\cite{Vinson08}. Accordingly, human subject data should be
confidential, benefits should outweigh real or potential risks and harms to
human data subjects or a society at large, and a study should be validly
conducted, having importance and relevance to software engineering as a
scientific discipline.

Due to the noted criterion of considering consequences for a majority, also the
relevance to software engineering should be extended to cover societal relevance
and impact, both of which are difficult concepts to exactly define. Roughly, the
former means that a research output should have a use and benefit to a society,
while the latter is concerned with a preferably measurable contribution to a
society, including its economy, governance, environment, culture, organizations,
or citizens~\cite{Bornmann13, Lindgreen21}. Within engineering sciences,
including software engineering, such uses, benefits, and contributions connote
particularly with large-scale, society-wide socio-technical software and other
systems~\cite{Polojarvi23, Ruohonen24REFSQ}. Such systems align well also with
the DSA's systemic risks and societal scope. Alternatively, however, the uses,
benefits, and contributions can also arise from smaller but particularly
relevant and impactful projects. To this end, the contexts of the scenarios
considered address a lack of rigorous definitions.

\subsubsection{Virtue Ethics}

Virtue ethics is concerned with a person's virtues, or his or her moral
character~\cite{Aristotle76}. Instead of rules, consequences, and acts,
ethics are approached by relating these to a person's character, such as his or
her honesty, fairness, compassion, empathy, and generosity. A common test in
virtue ethics involves asking whether an act will make you a better or a worse
person~\cite{MoutonKimppa15}. A truly virtuous person would know the answer
every time. When considering larger collectives, virtue ethics face the same
problems as deontology in a sense that people may agree upon neither virtues nor
rules. Therefore, a narrower branch of virtue ethics, professional ethics, is
often used in applied ethics research. Many professions have their own ethical
guidelines, often codified into so-called codes of conduct. While physicians and
medical practitioners would be a good example, lawyers and law practitioners
would be another.

\begin{table}[th!b]
\caption{Ethics of Persuasive Technology Design
  \cite[p.~52]{Berdichevsky99}}
\label{tab: eptd}
\begin{tabularx}{\linewidth}{lX}
\hline Principle & Description \\ \hline 1. & ``\small{The intended outcome of
  any persuasive technology should never be one that would be deemed unethical
  if the persuasion were undertaken without the technology or if the outcome occurred independently of persuasion.}'' \\
\cmidrule{2-2}
2. & ``\small{The motivations behind the creation of a persuasive technology should never be such that they would be deemed unethical if they led to more traditional persuasion.}'' \\
\cmidrule{2-2}
3. & ``\small{The creators of a persuasive technology must consider, contend with, and assume responsibility for all reasonably predictable outcomes of its use.}'' \\
\cmidrule{2-2}
4. & ``\small{The creators of a persuasive technology must ensure that it regards the privacy of users with at least as much respect as they regard their own privacy.}'' \\
\cmidrule{2-2}
5. & ``\small{Persuasive technologies relaying personal information about a user to a third party must be closely scrutinized for privacy concerns.}'' \\
\cmidrule{2-2}
6. & ``\small{The creators of a persuasive technology should disclose their motivations, methods, and intended outcomes, except when such disclosure would significantly undermine an otherwise ethical goal.}'' \\
\cmidrule{2-2}
7. & ``\small{Persuasive technologies must not misinform in order to achieve their persuasive end.}'' \\
\cmidrule{2-2}
8. & ``\small{The creators of a persuasive technology should never seek to persuade a person or persons of something they themselves would not consent to be persuaded to do.}'' \\
\hline
\end{tabularx}
\end{table}

As has been common~\cite{Anderson93, MoutonKimppa15, Winter18}, virtue ethics
are thus framed against the ACM's~\cite{ACM18} and IEEE's~\cite{IEEE14} codes of
conduct for computing professionals, including software engineers. In addition
to general ethical principles, both contain also responsibility mandates,
including a compliance with laws and regulations. In general, the ACM's code of
conduct is much more complete than the IEEE's one, but both are well-known and
well-accepted in computer science and software engineering as scientific
disciplines. However, like with many professional ethical guidelines, it is
still an open question whether and how well the two codes of conduct have been
adopted by professionals working in the software industry~\cite{Gogoll21}. Such
skepticism reflects notions such as ethics washing and the wider debates
surrounding ethics, self-regulation, and governmental regulation of software and
technologies in general~\cite{Green21, Ruohonen24FM}. Regardless, the ACM's and
IEEE's codes of conduct are useful for both research purposes and practical
applications. In addition, the virtue (or professional) ethics evaluation is
carried out with specific guidelines for the ethics of persuasive technology
design (EPTD). These guidelines are summarized in Table~\ref{tab: eptd}. Many of
these principles align with the ACM's and IEEE's codes of conduct. Regarding a
couple of specific principles worth pointing out, the seventh principle
considers deception, and the eight principle restates the earlier points about
consenting to persuasion.

\subsection{Scenario Analysis}\label{subsec: scenario analysis}

The analysis is based on imaginary scenarios, which have been typical in applied
ethics as a discipline, including its teaching in higher education~\cite{Brey12,
  Riser96}. In addition to analytical and pedagogical benefits, scenario
analysis can answer to the calls to move from essentially theoretical applied
ethics to contextual ethics and empirical ethics~\cite{Hoffmaster16}. As the
scenarios are imaginary, a step to empirical ethics is neither satisfied nor
possible, but contextual ethics provides the guideline; each scenario is
presented in a particular context that tries to mimic real-world settings as
closely as possible. Furthermore, the main concepts, deception and manipulation,
are considered not only in terms of dark patterns but also regarding software
engineering practices, including both in science and in industry. This duality
raises the practical relevance.

Regarding software engineering, scenario analysis has been popular within
requirements engineering, software architectures, usability, and some other
subdomains. Oftentimes, the overall rationale has been to improve or validate
plans and designs against desired functionalities~\cite{Dzida98}. Software
engineering provides also a further framing for the contextual analysis; most of
the scenarios considered follow the so-called technology transfer
model~\cite{Wohlin00}, considering university-industry software engineering
\text{projects---or}, rather, when considering the earlier framing to societal
relevance, societal impact, and large socio-technical systems,
university-industry-society projects. This threefold collaborative nexus is
particularly well-suited to approach various relevant ethical questions. It also
avoids a common pitfall in applied ethics addressing technology; the rigid
concentration only on a given technological artifact~\cite{Green21}.

To increase the practical relevance further, all scenarios presented involve two
evolutionary steps, thus analytically resembling a longitudinal research
setup. This type of a scenario analysis has been common in fields contemplating,
evaluating, and even predicting the emergence of future technologies and their
hypothesized consequences~\cite{Huss88, Wright14}. The approach and the fields
in general align with a concept of anticipatory ethics~\cite{Brey12}. Although
the present work is not about the future as such, the evolutionary steps
considered align well with the general wisdom in software engineering; among
other things, many software products are continuously changing and growing in
size~\cite{Ruohonen15JSEP}. Thus, as functionalities and deployment settings
evolve and change, so do the ethical questions involved. The point is also
important in terms of the deontological evaluation. As regulations and their
interpretations change through amendments and court cases, so do the
requirements drawn from the regulations~\cite{Hjerppe19RE}. In general, the
evolutionary focus raises the practical relevance because it allows to
contemplate changing ethical scenarios arising particularly from the technology
transfer model.

As for the practicalities, including structuring and presentation, each scenario
is approached and presented in three consecutive steps. The first step sets the
scene, outlining the case to consider and its specific context. The second step
presents an ethical evaluation of a scenario. The last step then concludes.

A simple but useful linear framework presented in existing research is used for
the analytical analysis~\citep{Riser96}. There are three steps also in the
framework. As is usual~\cite{Rashid09}, the first step is about finding the
facts and stakeholders involved. The second step is about making a defensible
ethical decision. This step involves isolating the core ethical issues,
examining potential legal issues, consulting guidelines, contemplating about the
three applied ethics branches, and finally making a decision that should be
defensible by sound argumentation. The last step is about providing instructions
to solve the situation in a wider practical context. These instructions should
involve listing the options available, enumerating recommendations and defenses
for these, and providing a short-term corrective measure based on the options
and recommendations. Although the framework was presented in a context of
teaching ethics in computer science, it contains the basic ingredients
applicable also for the present purposes. Finally, a note about defensible
claims: as there are seldom conjectures that could be formally proved in ethics,
a provision of a defense is about argumentative ethics. In terms of ethics of
argumentation, in turn, it is assumed that argumentative but imaginary
adversariality exists~\citep{Aikin21}. The imaginary adversaries who are seen to
be arguing are the three principal branches of applied ethics.

\section{Scenarios}\label{sec: scenarios}

\subsection{Privacy Research}

\subsubsection{Scenario}\label{subsec: privacy research}

\textit{Phase 1.} A European software engineering group partnered with privacy
researchers in the United States for evaluating how aware university students
are about dark patterns that try to manipulate people to disclose unnecessary
amounts of personal data. The privacy group designed a pattern and a technical
infrastructure for this task. Confidentiality of the personal data involved was
an explicit design goal. The European group then tested the pattern with about
thirty university students participating in a focus group. To raise the societal
relevance and impact, the two groups also partnered with a global media company
for launching a public awareness campaign based on the scientific
results. Before the actual research, a university's ethics council was
consulted, and all students participating to the focus group had to sign
handwritten forms for both scientific and data protection consents. A~debriefing
was also conducted to the students at the start of the focus group
session. \textit{Phase 2.} When preparing a publication based on the personal
data obtained in the focus group, which showed that awareness was rather low,
the European software engineering group received an email from the university's
administration: two students had complained that the group had allegedly
breached their privacy rights. Given the careful preparations, the complaint was
easily handled and no further examinations were conducted by the administration.

\subsubsection{Analysis and Verdict}

The scenario presents something that may happen in many human subject research
projects. In terms of dark patterns, the scenario deals with specific privacy
dark patterns, among which so-called ``privacy zuckering'' has been
common~\cite{Bosch16, Gray18, Gunawan22}. In general, this practice violates the
classical design principles about sound default
settings~\cite{Ruohonen25JISA}. Regardless, in terms of utilitarianism, the
scenario is ethical; raising awareness about privacy dark patterns has societal
benefits for the majorities in societies, especially when considering that the
preliminary research results indicated only low existing awareness. The
involvement of a media company in the awareness campaign planned indicates that
the two groups had also considered utilitarianism and consequences explicitly.

\begin{table}[th!b]
\caption{Was the Scenario in Section~\ref{subsec: privacy research} Ethical?}
\label{tab: privacy research}
\begin{tabular}{lccc}
\hline
& Deontology & Utilitarianism & Virtue ethics \\
\hline
Phase 1. & Yes & Yes & Yes \\
Phase 2. & Yes & Yes & Yes \\
\hline
\end{tabular}
\end{table}

As can be seen from Table~\ref{tab: privacy research}, the scenario is ethical
also in terms of virtue ethics. Although both the ACM's and IEEE's codes of
conduct emphasize protection of privacy, as do the principles in Table~\ref{tab:
  eptd}, the codes also emphasize confidentiality, including the protection of
personal data. The same applies to many other professional ethics
guidelines~\cite{Rothstein15}. The scenario description also makes it clear that
confidentiality was specifically taken into account when designing the privacy
dark pattern. The conclusion can be done by balancing the confidentiality
consideration against deontology and utilitarianism. The latter was already
considered, and there seems to be no ethical problems in terms of deontology
either. In fact, two consents were obtained from the student participants, a
debriefing session was held, and a university's ethics council was consulted
too. Although there may have still been problems with the conditions of
comprehension and competence, it was emphasized in Section~\ref{subsec:
  deception} that all conditions for informed consent are assumed to be valid
for the scenarios. Even though the scenario description does not allow to deduce
whether there might have been other problems with the GDPR, it can be concluded
that deontological considerations were satisfied also in this regard. Thus, all
in all, the scenario passes without any major remarks about ethics.

\subsection{Medical Applications}\label{sec: medical applications}

\subsubsection{Scenario}\label{subsec: scenario medical applications}

\textit{Phase 1.} A group of physicians at a big European hospital partnered
with software engineers working in the software industry for developing a new
mobile application in the domain of medical advise for people having a
diabetes. While all data protection requirements were carefully satisfied by
relying on the GDPR's exemptions for scientific research, the physicians lacked
technical knowledge to effectively postulate requirements for the software
development company involved. Despite this collaborative problem, the project
was a success. The physicians published two scientific articles on the mobile
application and the company involved later obtained all rights for a commercial
deployment. \textit{Phase 2.} Without further modifications, the application was
launched in Europe by distributing it on all mobile application ecosystems. An
aggressive marketing campaign helped the product launch. After a couple of years
in operation, the physicians started to receive complaints from people having a
diabetes. An investigation was conducted, and it was discovered that the
application only recommended costly diabetes drugs from particular foreign
pharmaceutical companies. Later on, it was further discovered that the company
had already before the commercial deployment developed an algorithm for
promoting drugs from particular companies and restricting the visibility of
cheaper drugs from other companies. It had also partnered with the said
companies already during the first phase. Alarmed by the results of the
investigations, the physicians retracted their articles and contacted a national
regulatory body.

\subsubsection{Analysis and Verdict}

The scenario, which resembles real-world cases~\cite{Powles17}, involves serious
ethical concerns. As the paper is about software engineering, the focus is on
the software development company and its engineers, not on the physicians who,
as said, have their own professional ethical principles. In any case, the first
concern is about data protection. During the first phase, all data protection
requirements were carefully taken account. In terms of deontology, the phase was
thus ethical in this regard. In the second phase, however, it was noted that the
mobile application was commercially launched without any
modifications. Therefore, the GDPR's exemptions for scientific research could
not have applied any longer; the commercial launch would have required choosing
a different legal basis for processing. A~violation is present also in terms of
the GDPR's purpose limitation.\footnote{~Article 5(1)(b) in Regulation (EU)
2016/679.}  In addition, personal health data belongs to the GDPR's category of
sensitive data. As was discussed in Section~\ref{subsec: deception}, processing
such data requires additional care. Likely, the company would have needed a
separate consent for this task. Thus, all in all, the second phase was unethical
according to deontology.

\begin{table}[th!b]
\caption{Was the Scenario in Section~\ref{subsec: scenario medical applications} Ethical?}
\label{tab: medical applications}
\begin{tabular}{lccc}
\hline
& Deontology & Utilitarianism & Virtue ethics \\
\hline
Phase 1. & No & Yes & No \\
Phase 2. & No & No & No \\
\hline
\end{tabular}
\end{table}

The second ethical concern relates to the manipulative algorithm designed and
implemented for promoting drugs from particular companies. It can be deduced
from the scenario description that the algorithm also limited the choices
available for users of the application. As was discussed in Section~\ref{subsec:
  regulations}, this dark pattern technique is the ease case for a legal
interpretation. Even without deontology, limiting the choices available is
typically seen as unethical because it violates an individual's
autonomy~\cite{Rothstein15}. Also some of the other laws may have been violated,
including the business-to-consumer directive. Furthermore, the scenario
description makes it clear that the company had designed and implemented the
algorithm already before the commercial launch. Although mobile applications are
not in the DSA's scope, the scenario reflects the issues about designing dark
patterns that were discussed earlier in Section~\ref{subsec: regulations}. For
these reasons, both phases are categorized as being unethical in term of
deontology.

Both phases are also unethical in terms of virtue ethics. As was already noted,
the ACM's and IEEE's codes of conduct emphasize compliance with laws, and thus a
clear violation of this ethical requirement is present in terms of the problems
with the GDPR. Furthermore, both codes mandate declaring any and all real or
potential conflicts of interest. The sixth principle in the EPTD conveys the
same message. As the software development company had partnered with the
pharmaceutical companies already during the first phase, it is clear that the
software engineers involved deceived not only the users of the application with
the deceptive algorithm but also their partnering physicians. This dual
deception is the third serious ethical concern. In general, it is against the
virtue of honesty \cite{Dougherty24}, which should be seen to be about not only
speaking the truth but also about being sincere and not
deceptive~\cite{Guenin05}. The conclusion of deceiving could be strengthened by
considering also the requirement of honesty in the ACM's code of conduct and the
requirement of respectfulness in the IEEE's code. As the physicians had to
retract their scientific articles, the deception also caused real harm to them.

Finally, utilitarianism might label both phases as ethical because helping
people having a diabetes is an ethical goal for the people and a society in
general. However, after a second thought, the second phase can be seen as
unethical also according to utilitarianism because financial hindrances were
caused to the people by promoting expensive drugs. In Europe free and fair
commercial competition is also a societal value. By majority voting, the verdict
is that the scenario is unethical all in all.

\subsection{Educational Platforms}\label{sec: educational platforms}

\subsubsection{Scenario}\label{subsec: educational platforms}

\textit{Phase 1.} Motivated by the successes of mobile applications and online
platforms for learning foreign natural languages, a group of university-based
software engineers in Japan won a substantial governmental research grant
together with a local software development company for developing and
researching a new platform for algorithmically aided free-of-charge
self-learning of Japanese. Explicitly or implicitly, they motivated the platform
by deception in the sense that Japanese was successfully framed and marketed,
including during the grant application, as an easy language to learn for
Europeans living in Japan. The research group and the company also implemented
and researched various manipulative techniques based on gamification that used
the personal learning data of the platform's users. A permission was
successfully obtained from a university's ethics council for testing the
platform with human subjects who were all Europeans. The project was a success
in terms of research; several articles were published in prestigious journals
that all proved the platform's effective capabilities for learning
Japanese. \textit{Phase 2.}  However, financially the project did not reach the
societal impact sought. As the company involved was unsuccessful at monetizing
the platform and lacked resources for maintaining it, the group and the company,
who jointly hold intellectual property rights, sold the platform to an Italian
university who stared to use it for teaching Japanese to its students. The deal
involved a promise that the platform would be fully compliant with Italian
laws. Yet, soon after the deal was made, the university's staff discovered that
the platform lacked all standard data protection requirements. After failing to
renegotiate or cancel the deal, the staff spent several months in fixing the
platform to comply with the requirements. Thereafter, the platform was
successfully used for providing self-learning of Japanese to students enrolled
to the university. Based on the data collected, the associated Italian
researchers concluded that Japanese was not an easy language to self-learn. They
published the finding in a prestigious scientific journal.

\subsubsection{Analysis and Verdict}

There are some ethical issues underneath the scenario, although only a few of
these seem to be major. To again start from deontology, the GDPR is not a
problem. Although the GDPR is an extraterritorial law and the legal
interpretations are complex~\cite{Gstrein21}, the regulation only applies to
processing of personal data of people within the~EU, regardless whether a
processor is within the union or outside of it.\footnote{~Article 3 and recitals
from 22 to 25 in Regulation (EU) 2016/679.} Although the scenario description
makes it clear that the online language learning platform was specifically
offered to and tested with Europeans living in Japan, the GDPR, unlike criminal
law in some extraterritorial contexts~\cite{DeHert18, Pocar04}, does not apply
in this case. Therefore, also the missing of data protection requirements is not
an issue from a deontological perspective. As these were later on implemented by
the Italians, the same conclusion about deontology applies to the second phase.

\begin{table}[th!b]
\caption{Was the Scenario in Section~\ref{subsec: educational platforms} Ethical?}
\label{tab: educational platform}
\begin{tabular}{lccc}
\hline
& Deontology & Utilitarianism & Virtue ethics \\
\hline
Phase 1. & Yes & Yes & Yes \\
Phase 2. & Yes & Yes & No \\
\hline
\end{tabular}
\end{table}

However, the selling of the leaning platform seems to have violated a contract
between the two parties as full compliance was promised by the sellers. As the
scenario description does not make it clear whether the buyers might further
pursue the case in a court, a definite answer about a breach of a contract
cannot be given. Also due to this uncertainty, the second phase is
labeled as ethical in terms of deontology in Table~\ref{tab: educational
  platform}. Furthermore: although dark patterns were implemented, these seem to
be beyond the reach of the EU laws. The scenario description notes that the
platform was a free-of-charge service during the first phase. Nor can its use in
a university context be considered as a commercial activity in the second
phase. The platform seems to also be beyond the DSA's scope because its user
base in the scenario is presumably only modest. Vetting is also present in the
second phase because only students enrolled to a university were among the
users.

No ethical problems are present in terms of utilitarianism. Helping people to
learn new languages is a noble and laudable goal. It benefited the Japanese
society, the Europeans living in Japan, and the Italian students. These points
align with the usually utilitarian evaluation of the ethics of gamification in
general~\cite{Marczewski17, Zvereva23}. Having said, there may be wider societal
concerns involved; the platform may have contributed to increasing digital
addiction among students, which would have negative consequences for them and
the two societies in general. As such potential consequences are open to a
debate, and no clear and universally accepted ethical or legal guidelines
exist~\cite{Berthon19}, the initial conclusion still seems adequate and
justified.

The scenario description raises alarms in terms of virtue ethics particularly
during the second phase. There might be ethical issues also during the first
phase, but the issue seems to be really about anticipatory ethics, about
deducing whether the lack of compliance with the GDPR was known beforehand,
before selling the platform. If so, and if they lacked legal knowledge, they
should have consulted legal experts, as also pinpointed by a requirement about
performing work only in areas of competence in the ACM's code of conduct. The
same point applies to the Japanese ethics council from whom the software
engineers successfully gained a permission to test the platform with
Europeans. If it is assumed that they knew about the GDPR's requirements, then
the advertising of full legal compliance is clearly also lying, and in this case
the deception can be seen as serious enough to violate also virtue ethics,
including those drawn from the codes of conduct for computing professionals. Due
to the apparent uncertainty, the first phase can be seen to be on the ethical
side, or maybe the gray zone, but the breach of a contract seems serious enough
to label the second phase as unethical in terms of virtue ethics---even in case
the breach was unintentional.

Finally, some perhaps minor issues were present also during the grant
application, given that the scenario description hints that they knew that
Japanese was not an easy language to learn for Europeans. Though, it remains
debatable whether this form of deception is more on the side of
marketing. Although the Italian researchers refuted the findings of the Japanese
researchers, also the latter published their findings in good journals, and thus
the remaining issues involved are a matter of a scientific debate and not virtue
ethics \textit{per~se}. All in all, majority voting concludes about overall
ethical conformance.

\subsection{Phishing}\label{sec: phishing}

\subsubsection{Scenario}\label{subsec: phishing}

\textit{Phase 1.} A software engineering department in a big European university
received a task from the university's administration to develop and test
security awareness campaigns. The rationale was to improve the resilience of the
university against phishing attacks, which had become a big problem in recent
years. The department's software engineers designed the campaigns around
email. In three consecutive campaigns, the university's staff was targeted with
non-malicious phishing emails and statistical observations were collected on
those staff members who had fallen to the phishing emails. The university's
staff was debriefed about the campaigns beforehand. The data collection was
fully anonymized. \textit{Phase 2.} The campaigns were successful: security
awareness increased substantially. Nevertheless, the administration was not
satisfied, as it believed that the campaigns were too simplified for emulating
the real phishing attacks the university had witnessed. To this end, the
department received another request to develop a better campaign. The software
engineers used two techniques for this task. They first trained a large language
model (LLM) on the staff's email correspondences, and then designed a phishing
website that replicated the university's real website. Then, the personalized
LLM-based phishing emails were again delivered to the staff. In addition to
again anonymously recording statistics on the amount of staff members who
clicked links in the phishing emails, they anonymously recorded statistics on
those staff members who further used the login functionality of the phishing
website to which the links pointed to. According to the results, a substantial
amount of staff members fell victim to this simulated security awareness
campaign. After asking a permission from the administration, the software
engineers involved published the findings of the four campaigns in a scientific
journal.

\subsubsection{Analysis and Verdict}

The scenario continues the motivating discussion in the introduction about the
close relationship between phishing and dark patterns. To start from
utilitarianism, the scenario can be seen as ethical during both phases. As
security awareness increased among staff, even though the last campaign
indicated further work to be done, the university and staff both clearly
benefited from the work done. In other words, the requirement of beneficence was
satisfied. As the use of LLMs in (spear) phishing research is a new and intriguing
research area~\cite{Bethany24}, the scientific publication can be also assumed
to have had scientific value. The scenario's ethical problems are elsewhere.

\begin{table}[th!b]
\caption{Was the Scenario in Section~\ref{subsec: phishing} Ethical?}
\label{tab: phishing}
\begin{tabular}{lccc}
\hline
& Deontology & Utilitarianism & Virtue ethics \\
\hline
Phase 1. & Yes & Yes & Yes \\
Phase 2. & No & Yes & No \\
\hline
\end{tabular}
\end{table}

In terms of both deontology and virtue ethics, data protection is a concern but
not during the first phase; as the scenario description makes it clear,
anonymization was done and the staff was also debriefed in advance, hinting that
the campaigns perhaps followed the research-based deception techniques noted in
Section~\ref{subsec: deception}. The problem is located in the second
phase. Specifically, it is about the training of the LLM with the staff's
personal emails. Although analogous things have been commonly done in the
so-called insider threat research~\cite{Ruohonen24ICDF2C}, these are problematic
in terms of data protection. In addition to the GDPR, many European countries
have established specific data protection rules in the domain of labor law. As
the scenario description does not make it clear in which European country the
university was located, a detailed analysis is impossible but it can be
concluded with a reasonable confidence that the training was problematic
legally. They should have likely obtained informed consents from the
staff~\cite{Resnik18}, perhaps doing also privacy impact
assessments. Furthermore, the apparent lack of autonomy and agency is
problematic from a deontological perspective because without these people may
start treating others not as ends of themselves but as means, thus violating the
Kant's second formulation of the categorical imperative.  Analogously to the
data protection laws in Europe, including the GDPR, the principles in
Table~\ref{tab: eptd} as well as the ACM's code of conduct also generally
emphasize transparency, informed consent, purpose limitation, and many related
things related to privacy. Although the scenario's second phase can be concluded
to be unethical, the problems could have been fixed with a relatively little
amount of effort.

\subsection{Citizen Platforms}\label{sec: citizen platforms}

\subsubsection{Scenario}\label{subsec: citizen platforms}

\textit{Phase 1.} The government of Germany approached a prestigious German
university for developing a participatory citizen platform to help at
deliberative democracy. Thanks to the generous funding received, the software
engineers working at the university successfully designed and implemented such a
platform. Although not named as such, various algorithmic manipulation
techniques were designed to help the deliberation and its monitoring. The
platform was then successfully piloted in a context of urban planning and
associated policy-making involving a complete overhaul of a public transport
system in a major German city. An evaluation of the pilot was published in a
scientific journal. \textit{Phase 2.} Motivated by the success of the pilot, the
government deployed the platform on a wider scale, opening it to all Germans
wanting to discuss politics and policy-making in the country. However, after
about a year in operation, the civil servants associated with the platform
discovered that people had found a way to exploit the manipulative
algorithms. Disinformation and hate speech were now rampart. After consulting
the university, it was discovered that the software engineers involved with the
pilot study had faced similar issues, having had to devote substantial effort
for content moderation and monitoring of the algorithms. The civil servants then
had to devote a large amount of new financial resources for content moderation,
involving recruitment of several new employees. After about a year, it was
discovered that even more resources would be needed. By now, the problems had
been noticed also by media, and a public scandal endured. Pressured by
high-profile politicians, the civil servants had no option but to shutdown the
platform.

\subsubsection{Analysis and Verdict}

The scenario presents an intriguing short imaginary story on how
well-intentional manipulative algorithms can go to astray. Therefore, the
scenario also resembles the real-world cases involving large social media
platforms, the exploitation of algorithms that are allegedly exploitative
themselves, and the real or perceived new dangers posed by artificial
intelligence in these settings~\cite{Garon22}. It also serves as a warning for
software engineering because similar platforms have been recently envisioned
within the discipline~\citep{Ruohonen24REFSQ}. Having said that, ethically the
scenario seems to be only modestly problematic.

As the scenarios is about Germany, it can be started by noting that all GDPR
requirements were likely implemented, although the scenario description does not
make it explicit. Analogously to the language learning platform scenario in
Section~\ref{sec: educational platforms}, it seems that the DSA does not apply
during the first phase because the scope was limited. In the second phase,
however, the opening of the platform to all Germans seems to entail a sufficient
amount of potential people to satisfy the DSA's notions about potentially
unlimited recipients. Therefore, the second phase is labeled as unethical in
terms of deontology in Table~\ref{tab: citizen platforms}.

\begin{table}[th!b]
\caption{Was the Scenario in Section~\ref{subsec: citizen platforms} Ethical?}
\label{tab: citizen platforms}
\begin{tabular}{lccc}
\hline
& Deontology & Utilitarianism & Virtue ethics \\
\hline
Phase 1. & Yes & Yes & Yes \\
Phase 2. & No & No & No \\
\hline
\end{tabular}
\end{table}

When considering the software engineers involved, the second phase can be seen
as unethical also in terms of virtue ethics. Given that the scenario description
makes it clear that they knew about the problems, they should have at least
warned the civil servants beforehand. The ACM's code of conduct also emphasizes
that special care should be taken with software or other systems that are going
to be integrated into an infrastructure of a society. Predictable outcomes are
also emphasized in the third principle of the EPTD guidelines. While the ACM's
code also emphasizes due diligence to ensure that a system functions as
intended, the first phase can be seen as ethical because there are no easy
answers to content moderation issues, with or without the presence of
manipulative algorithms. It may be that the issues were also a mere oversight,
not an intention to deceive.

The first phase is ethical according utilitarianism. Improving democracy, civic
engagement, and deliberation is an important and ethical goal. The pilot was
also successful and supposedly improved the democratic planning of the new
public transport system. However, the second phase raises a question whether
consequences should be evaluated in terms of their intended or realized
consequences. This question leads to a long-standing and unresolved
philosophical debate between actual versus probable (or anticipatory)
utilitarianism~\cite{Gruzalski81, Strasser89}. The risk-based approach in many
EU regulations would perhaps provide a middle-ground. In other words, the
software engineers should have probably conducted risk assessments and delivered
these to the civil servants. As was noted in Section~\ref{subsec: deontology},
risks can be also seen as important both in terms of deontology and
utilitarianism. On these grounds, the second phase is also seen as unethical. In
terms of risks, a limitation in the DSA can be also remarked in that it
restricts the evaluation of systemic risks only to very large platforms and
search engines. As the scenario demonstrates, however, at least moderate
societal risks can arise also from much smaller software engineering
projects. In any case, majority voting again declares a draw.

\subsection{Honeypots}\label{sec: honeypots}

\subsubsection{Scenario}\label{subsec: honeypots}

\textit{Phase 1.} A software engineering group specialized into cyber security
took part to a large EU-funded consortium researching innovative means to
improve threat detection. The group's work package involved designing new
deception techniques for honeypots, which were based on introspection methods
for virtual machines. Given the large-scale nature of the consortium and the
funding body's requirements, ethical, legal, and many other factors were
considered to the finest detail, including with third-party legal
experts. \textit{Phase 2.} Although the group's role in the consortium was only
minor, it completed its work successfully. Pushed by the innovativeness of the
results obtained, the group quickly published its results as three so-called
pre-prints in order to gain a first-mover status, but without consulting the
consortium in advance. A company in the consortium also showed an interest to
develop the techniques further. However, soon after the pre-prints were
published, the group and the consortium noticed that an anonymous cyber criminal
group had already reverse engineered the techniques and published their findings
in the dark net. The company involved subsequently canceled its plans to engage
further with the honeypot techniques.

\subsubsection{Analysis and Verdict}

The scenario is highly relevant in terms of deception as a theoretical
concept. Honeypots are fundamentally about deception; they try to deceive cyber
criminals and other intruders into believing that a system they have compromised
or otherwise assessed is a real system, when, in fact, it is merely a technical
decoy. In a sense, honeypots are thus a ``grand dark pattern''. In terms of
scientific value, which, as was noted in Section~\ref{subsec: utilitarianism},
is an important part for an utilitarian evaluation in this scenario, honeypot
deception techniques are an active research domain~\cite{Javadpour24}. The
anticipatory consequence of improving cyber security is also ethical. Therefore,
the first phase can be seen as ethical in terms of utilitarianism. The second
phase raises the same problem as the earlier citizen platform scenario in
Section~\ref{sec: citizen platforms} with respect to anticipatory and actual
utilitarianism. As anti-honeypot and anti-introspection methods are
well-known~\cite{Uitto17}, the software engineering group probably knew, or at
least should have known, about the risks involved. Potential harm was also
caused to the company involved. Although these points allow to label the second
phase as unethical according to utilitarianism, the real problem is really about
a much bigger issue.

\begin{table}[th!b]
\caption{Was the Scenario in Section~\ref{subsec: honeypots} Ethical?}
\label{tab: honeypots}
\begin{tabular}{lccc}
\hline
& Deontology & Utilitarianism & Virtue ethics \\
\hline
Phase 1. & Yes & Yes & Yes \\
Phase 2. & No & No & No \\
\hline
\end{tabular}
\end{table}

The bigger issue is about academic freedom. There are several dimensions to
academic freedom, which is not an absolute right but must be balanced against
other conditions for scientific integrity, but in the present context it means a
freedom to research without undue interference, pressure, or limitation, whether
political, administrative, or other kind~\cite{EPRS24}. Now, in the present
scenario the question is whether the group should have exercised this right by
publishing the results even if they knew about the risks. The scenario
description makes it clear that the consortium had considered governance in
detail, and thus it can be assumed to have had also specific rules regarding
publication, as is generally recommended~\cite{Morrison20}. Thus, it can be
concluded that the software engineering group should have negotiated with the
company and the consortium in general before publishing the results, although,
even then, the same issue about academic freedom is present. The ACM's code of
conduct has also many related clauses, including a responsibility to seek and
accept peer and stakeholder reviews. For these reasons, the second phase is
labeled as unethical in terms of both virtue ethics and deontology. All in all,
given Table~\ref{tab: honeypots}, the publication and negotiation mishaps make the
scenario again a draw according to majority voting.

\section{Discussion}\label{sec: discussion}

\subsection{Conclusion}

The conclusion can be summarized with five points:

\begin{enumerate}
\item{Data protection and the GDPR can be easily used to postulate issues that
  are problematic in terms of deontology, virtue ethics, and research ethics in
  general. The examples considered included sensitive data and changing legal
  basis of processing due to evolution in a university-industry software
  engineering project (Section~\ref{sec: medical applications}),
  extraterritoriality in a similar project (Section~\ref{sec: educational
    platforms}), and processing of personal data of employees (Section~\ref{sec:
    phishing}). It seems reasonable to claim that these problems may be common
  also in practice.}
\item{The EU's laws explicitly or implicitly addressing dark patterns are a
  little harder to evaluate, but reasonably well-contextualized scenarios were
  also postulated for these. The examples included a case that likely violated
  the business-to-consumer directive and the DSA due to a promotion of
  particular products and restricting the visibility of others
  (Section~\ref{sec: medical applications}) and opening a citizen platform with
  manipulative algorithms to a large group of people (Section~\ref{sec: citizen
    platforms}). The former was about a university-industry project, while the
  latter was on the side of university-industry-society projects.}
\item{Relevant problems were demonstrated also in terms of utilitarian
  ethics. The examples included financial harms caused to people suffering from
  a medical illness (Section~\ref{sec: medical applications}), causing harm to a
  whole society in terms of deliberative democracy due to a lack of proper
  risk-analysis in a software engineering project (Section~\ref{sec: citizen
    platforms}), and causing harm to an industry partner in a large
  university-industry software engineering project (Section~\ref{sec:
    honeypots}). If the scenario analysis would have concentrated strongly on
  anticipatory utilitarianism, the list of problems would have been larger.}
\item{Many relevant problems were elaborated also in terms of virtue ethics and
  its narrower cousin, professional ethics. By using the ACM's code of conduct
  as a reference point, the examples included violations of the mandates to
  respect laws and rules (Section~\ref{sec: medical applications}, among
  others), declare conflicts of interest (Section~\ref{sec: medical
    applications}), conduct work only on one's areas of competence
  (Section~\ref{sec: educational platforms}), respect privacy and data
  protection (Section~\ref{sec: phishing}), taking special care with societal
  projects (Section~\ref{sec: citizen platforms}), and seeking peer and
  stakeholder reviews (Section~\ref{sec: honeypots}). Many of these align with
  the deontological problems elaborated and discussed.}
\item{Many of the scenarios considered dealt with anticipatory ethics. The
  scenario analysis presented thus further demonstrates how ethical questions
  often change over time. Anticipating the future is important also in
  ethics. To this end, it may make sense to align ethical evaluations with other
  anticipatory techniques, such as risk analysis that is common in cyber
  security. Deontology is involved too because many laws, including the GDPR,
  mandate risk and impact assessments.}
\end{enumerate}

It can be also concluded that particularly the ACM's code of conduct is useful
for virtue ethics evaluations. It is sufficiently comprehensive yet flexible to
deduce about various potential ethical issues involved in software engineering
projects, including those related to dark patterns. However, it is not
sufficient alone because also deontology and utilitarianism should be considered
in an evaluation. To this end, as has been argued also
previously~\cite{MoutonKimppa15}, the two-step evolutionary scenario analysis
with the three main branches of normative ethics may be useful for both
ethics councils in universities and teaching of applied ethics in software
engineering.

\subsection{Limitations}

Three limitations should be briefly acknowledged. First, argumentative ethics
and applied ethics in general necessarily include a degree of
subjectivity. Although subjectivity was decreased and objectivity increased by a
mandate that all authors of the paper had to agree with the results from each
scenario, objectivity or its transparency could have been further enhanced by
recording also the disagreements and negotiations between the authors. It would
have raised the paper's validity in a sense that appealing to a consensus in
normative questions is not without problems~\cite{Davies15}. In other words, it
would have been possible to use inter-rater reliability measures sometimes used
in systematic literature reviews and qualitative analysis involving coding of
patterns, themes, or other constructs.

Second, despite the careful contextualization, it remains unclear how
representative the theoretical scenarios are with respect to ethical problems
encountered in real software engineering projects. The third limitation is
closely related: although a move from theoretical applied ethics to contextual
ethics was made, a leap to empirical ethics was not taken. Addressing this
limitation would be a good and promising way to continue in further research. An
empirical evaluation would be needed also regarding the earlier note about the
alleged usefulness and pedagogical value of the scenario analysis approach
presented.

\subsection{Further Work}

The paper opens a door for many promising paths in further ethics research. On
one hand, first, further empirical research is required on the ethical problems
theoretically but contextually outlined. Such research should focus also on the
fundamental concepts involved, including deception, manipulation, and individual
autonomy. Analogously to AI ethics \cite{Whittlestone19}, it seems reasonable to
assume that different tensions exist among software engineers about the meaning
of the concepts and their real or perceived consequences to people using
software containing dark patterns. As was noted in Section~\ref{subsec:
  manipulation}, particularly the concept of manipulation is difficult to
define.

Empirical ethics would allow to also move beyond the common individualist
foundation in ethics dealing with technology. In this regard, it has been
forcefully argued that it may not matter much whether software and other
engineers are truly virtuous persons, trying to comply with all regulations and
always following professional codes of conduct, in case there is no support from
a management~\cite{Green21}. To this end, it might make more sense to study
ethics of collectives, such as software engineering research groups, software
engineering teams in the software industry, and particularly managers of
software development companies. A related group for empirical ethics research
would be the people forced to use software with dark patterns. Although there is
existing research on people's perceptions and opinions about dark patterns,
their ethical concerns about these have not been addressed specifically.

On the other hand, second, also further theoretical research is needed. Both
conceptual and legal approaches would be required to better understand the
fundamental concepts involved. While deception has been fairly well theorized in
the literature, many questions remain about manipulation as a concept. The
problem is already well-recognized in philosophical research~\cite{Christiano22,
  Jongepier22}. In particular, manipulation as a concept should be conceptually
and theoretically compared to other related concepts, such as power and
influence. In fact, the definition that was given in the opening of
Section~\ref{subsec: manipulation} is almost identical to a classical definition
of power in political science~\cite{Dahl57}. It is also identical to definitions
given for interpersonal persuasion~\cite{Fogg08}. Gamification would be a
further example. Thus, a better understanding is required on whether and how
manipulation allegedly differs from the other concepts. Similar points apply to
the other fundamental concepts involved, including individual autonomy and harm
in particular. These concepts are fundamental also in terms of
ethics. Understanding real or potential harms caused is essential for
utilitarianism, while a loss of autonomy strikes at the heart of ethics. If one
has no autonomy, one cannot reason ethically either.

Third, the conceptual problems affect also the EU's laws explicitly or
implicitly addressing dark patterns. The associated case law too seems rather
difficult to deduce about regarding the concepts~\cite{Trzaskowski24}. To this
end, a close eye is needed to keep track on whether the regulators or courts
will come up with a clear precedent in the future. Now that there are ongoing
DSA investigations on some online marketplaces where dark patterns are
plentiful~\cite{EC24}, such a precedent may also emerge in the nearby future.

Fourth, it seems reasonable to recommend that further research is needed on the
actual technical implementations behind dark patterns. As was noted in
Section~\ref{subsec: regulations}, these are likely important also in terms of
law. The point reiterates earlier arguments raised about a lack of technical
investigations in the evaluation of design approaches and their
ethics~\cite{Gerdes23}. From a broader perspective, the question is about how to
incorporate moral values into software.

Fifth, theoretical and conceptual research is needed also regarding
the concept of dark patterns itself. The present paper diverged slightly from
the conventional understanding in that dark patterns were framed to cover
software beyond user interface designs. This choice can be justified on the
grounds that particularly the literature dealing with democracy, politics, and
algorithmic manipulation is concerned with many other technological
artifacts. Clarifications are required also on how such artifacts are related to
user interface designs. If the designs feed on personal data and machine
learning algorithms, like many of them today do, conceptual work is required on
the meaning of the whole concept of a user interface. As was discussed, similar
theoretical points apply to artifacts in the domain of cyber security, such as
phishing websites and honeypots.

Last, it would be worthwhile to consider, evaluate, and extend ethical
evaluations in software engineering more generally. As was seen, the three
normative branches are often in a conflict with each other. Such conflicts
create challenges for final decisions about right or wrong. To resolve conflicts
and to secure just relations between different, possibly opposing groups, a
dialogue and a rational discourse are often seen as essential. To this end, a
branch known as Habermasian discourse ethics~\cite{Habermas90, Habermas93} has
emerged as a promising philosophical framework, including in the context of
computing and technologies~\text{\citep{Stahl12, Yetim06}}. New ideas have also
been proposed for using non-commercial technology platforms for deliberative
democracy and beyond, including for requirements
engineering~\cite{Ruohonen24REFSQ}. Such ideas together with discourse ethics
could also shed further light on the ethics of dark patterns. Epistemologically
sound argumentative discussion could also be a way to define manipulation
through a negation. More generally, these would offer tools to solve situations
in which either ethical theories or actors involved have different viewpoints
about ethics. In other words, discourse ethics offer a promising path to solve
conflicts and to address situations in which a conclusion is a
draw~\cite{Mingers10}. That said: if the main ethical theories implicate that
something is ethical or unethical, it likely also is.

\balance
\bibliographystyle{apalike}

\end{document}